\documentclass[fleqn,usenatbib]{mnras}

\usepackage{newtxtext,newtxmath}

\usepackage[T1]{fontenc}
\usepackage{ae,aecompl}

\usepackage{graphicx}	
\usepackage{amsmath}	
\usepackage{amssymb}	
\usepackage{pdflscape}

\usepackage[figure,figure*]{hypcap}
\usepackage{rotating}

 \usepackage{multirow}

\newcommand{\Hip}{{\emph{Hipparcos~}}}

\newcommand{\logg}{$\log g$}

\newcommand{\ds}{$\delta$ Sct}
\newcommand{\dss}{$\delta$ Sct stars}

\newcommand{\corot}{{CoRoT}}

\newcommand{\kepler}{{\it{Kepler}}}

\newcommand{\vsini}{$\mathrm{v}\cdot\!\sin\!~i$}

\newcommand{\msol}{$\mathrm{M}_{\odot}$}

\newcommand{\cd}{$\mbox{d}^{-1}$}

\newcommand{\Dnu}{$\Delta\nu$}

\newcommand{\rhom}{$\bar\rho$}

\newcommand{\gh}{Garc\'{\i}a Hern\'andez}

\newcommand{\stard}{HD\,172189}
\newcommand{\stare}{CID\,100866999}

\newcommand{\starg}{HD\,159561}
\newcommand{\starh}{KIC\,9851944}
\newcommand{\stari}{KIC\,8262223}
\newcommand{\starj}{KIC\,10080943}
\newcommand{\stark}{HD\,15082}

\title[Precise surface gravities of $\delta$ Scuti stars from asteroseismology]{Precise surface gravities of $\delta$ Scuti stars from asteroseismology}

\author[Garc\'{\i}a Hern\'andez et al.]{A. \gh$^{1,2}$,\thanks{agh@ugr.es}
J. C. Su\'arez$^{1,3}$,
A. Moya$^{1}$,
M\'ario J. P. F. G. Monteiro$^{2,4}$,
\newauthor
Z. Guo$^{5,6}$,
D. R. Reese$^{7}$,
J. Pascual-Granado$^{3}$,
S. Barcel\'o Forteza$^{8,9}$,
S. Mart\'{\i}n-Ruiz$^{3}$,
\newauthor
R. Garrido$^{3}$,
J. Nieto$^{10}$
\\
$^{1}$University of Granada (UGR), Dept. Theoretical Physics and Cosmology, 18071, Granada, Spain\\
$^{2}$Instituto de Astrof\'{\i}sica e Ci\^{e}ncias do Espa\c{c}o, Universidade do Porto, CAUP, Rua das Estrelas, PT4150-762 Porto, Portugal\\
$^{3}$Instituto de Astrof\'{\i}sica de Andaluc\'{\i}a (CSIC), Glorieta de la Astronom\'{\i}a S/N, 18008, Granada, Spain\\
$^{4}$Departamento de F\'{\i}sica e Astronomia, Faculdade de Ci\^{e}ncias da Universidade do Porto, Rua do Campo Alegre, 4169-007 Porto, Portugal\\
$^{5}$Copernicus Astronomical Center, PAN, Bartycka 18, 00-716 Warsaw, Poland\\
$^{6}$Center for High Angular Resolution Astronomy and Department of Physics and Astronomy, Georgia State University, P. O. Box 5060,\\ Atlanta, GA 30302-5060, USA\\
$^{7}$LESIA, Observatoire de Paris, PSL Research University, CNRS, Sorbonne Universit\'es, UPMC Univ. Paris 06, Univ. Paris Diderot,\\ Sorbonne Paris Cit\'e, 5 place Jules Janssen, 92195 Meudon, France\\
$^{8}$Instituto de Astrof\'{\i}sica de Canarias, 38200 La Laguna, Tenerife, Spain\\
$^{9}$Departamento de Astrof\'{\i}sica, Universidad de La Laguna, 38206 La Laguna, Tenerife, Spain\\
$^{10}$Joan Rebull 92, 43850 Cambrils, Tarragona, Spain
}

\date{Accepted XXX. Received YYY; in original form ZZZ}

\pubyear{2017}

\begin{document}
\label{firstpage}
\pagerange{\pageref{firstpage}--\pageref{lastpage}}
\maketitle


\begin{abstract}
The work reported here demonstrates that it is possible to accurately determine surface gravities of \dss\ using the frequency content from high precision photometry and a measurement of the parallax. Using a sample of 10 eclipsing binary systems with a \ds\ component and the unique \ds\ star discovered with a transiting planet, WASP-33, we were able to refine the \Dnu-\rhom\ relation. Using this relation and parallaxes, we obtained independent values for the masses and radii, allowing us to calculate the surface gravities without any constraints from spectroscopic or binary analysis. A remarkably good agreement was found between our results and those published, extracted from the analysis of the radial velocities and light curves of the systems. This reinforces the potential of \Dnu\ as a valuable observable for \dss\ and settles the degeneracy problem for the \logg\ determination through spectroscopy.
\end{abstract}

\begin{keywords}
binaries: eclipsing --- stars: oscillations --- stars: rotation --- stars: variables: delta Scuti
\end{keywords}

\section{Introduction}

$\delta$ Scuti (\ds) stars are A to F-type pulsating stars located on the main sequence or at the H-shell burning phase. They are placed within the classical instability strip with a mass between $\sim$1.5 and 3 \msol. The interpretation of the oscillation spectra of \dss\ is not an easy task. The intrinsic nature of the modes and the interaction with other important mechanisms, namely rotation, complicates the oscillation spectrum \citep[see e.g.][ for an interesting review on the subject]{Goupil2005}. In the era of space missions like \emph{MOST} \citep{most} \corot\ \citep{corot} and \kepler\ \citep{kepler}, despite the significant increase in the number of detected frequencies, the frequency content is still hard to decipher \citep[e.g. ][]{poretti2009, garciahernandez2009, Javier2015}. Even so, relevant progress has been achieved regarding the theoretical modelling of these stars thanks to the inclusion of rotation effects in 2D evolution models, and the use of a non-perturbative approach to compute the oscillation frequencies and mode visibilities \citep[see e.g.] [to name a few]{Reese2013, Ouazzani2012, Rieutord2016}. Some advances have been shedding light on the interpretation of the pulsation spectrum thanks to the analysis of frequency patterns \citep{garciahernandez2009, garciahernandez2013, Paparo2016sample, Sebastia2017, Michel2017, Reese2017}.\par

A-type stars are hot and usually fast rotators. Measurements of the surface projected velocities \vsini\ and statistical inferences on the real rotation rates show that they can reach values close to the break-up limit \citep{Royer2007}, which severely hampers accurately determining stellar global parameters, such as effective temperature and abundances. In addition, the lack of metallic lines in the spectra is responsible for the well-known degeneracy in surface gravity determinations.\par

Binary systems with a \ds\ component can help with handling these problems. Accurate physical parameters can be obtained thus avoiding any degeneracy. Observing these systems with recent space satellites allowed us to properly resolve the pulsation spectra of their \ds\ components. In a recent work, \cite{gh2015} (GH15 in what follows) demonstrated that there exists a direct relation between the mean density of \dss\ and one of these frequency patterns, a large frequency separation, defined as the difference in frequency between pressure modes (p modes) of the same spherical degree, $\ell$, and consecutive radial orders, $n$.\par

In this work, we use a well studied sample of eclipsing binary systems with a \ds\ component to determine the stellar surface gravities only from the pulsation frequencies, which breaks the degeneracy for this physical quantity found by spectroscopic methods. This allows us to accurately measure the physical parameters for A-F type stars.\par

\section{The sample}
\label{sec-2}

The starting point was the sample of 7 eclipsing binaries in GH15. We added four additional well studied systems: three binary systems and a \ds\ star with a transiting planet. The three binary systems, \starh\ \citep{Guo2016}, \stari\ \citep{Guo2017} and \starj\ \citep{Schmid2015}, were observed by the \kepler\ satellite. The combined analysis of their light curves and radial velocity measurements allows us to determine the physical parameters for both components. \starh\ and \stari\ are eclipsing binary systems, but \starj\ is indeed an eccentric, non-eclipsing binary showing periastron brightening events. Nonetheless, a similar analysis to those of the eclipsing systems allows an accurate and precise determination of the masses and radii. Additionally, we included in our study \stark, also known as WASP-33, the only \ds\ star discovered with a transiting planet. For this object, the mean density is also measurable in a similar way to the other stars in the sample \citep{CollierCameron2010}.\par

The relevant parameters of these systems are listed in Table~\ref{table1}. The mean density (\rhom), the mean radii ($\bar{R}$), the surface gravity (\logg$_{b}$) and the large separation (\Dnu) have been obtained using the values found in the literature (see Sec.~\ref{sec-3} for details). We only derived \Dnu\ values for \starj\ and \stark\ for the first time.\par

In some cases, both components in the system showed pulsations and it was not possible to determine which one is responsible for the large frequency separation found. In our analysis, we did not find any signatures of multiple large separations. Thus, all the parameters in the table refer to the star whose properties best fits the large separation-mean density relation. In the unlikely case that the pattern was a combination of two large separations, it would not fit the \Dnu-\rhom\ relation.\par

Parallaxes ($\hat{\pi}$), needed to get an independent estimate of the masses were taken from the Gaia mission catalogue \citep{gaiamission,gaiaDR1} or from \Hip \citep{vanLeeuwen2007}. We describe the complete procedure on how to obtain masses and surface gravities from the parallaxes in Sec.~\ref{sec-4}. But we need to determine accurately mean densities first.\par

\begin{table*}
 \begin{footnotesize}
  \caption{Characteristics of the systems taken from the literature (see footnote for references) and calculated in this work. Columns 3 to 7 are values inferred from the binary analysis, whereas column 8 is calculated from the parallax. The information corresponds to the component showing the large frequency pattern (which is not necessarily the primary).}
  \label{table1}
   \begin{tabular}{llllllllllllllllll}
  \hline
  \hline
   \multicolumn{1}{c}{System} &
   \multicolumn{1}{c}{$\Delta\nu$} &
   \multicolumn{1}{c}{$\bar\rho$ } &
   \multicolumn{1}{c}{$\Omega/\Omega_{\mathrm{K}}$} &
   \multicolumn{1}{c}{$M_\mathrm{b}$} &
   \multicolumn{1}{c}{$\bar{R}$} &
   \multicolumn{1}{c}{$\log g_\mathrm{b}$} &
   \multicolumn{1}{c}{$\log g_\mathrm{\hat{\pi}}$} &
   \multicolumn{1}{c}{$\hat{\pi}$} \\
   \multicolumn{1}{c}{ } &
   \multicolumn{1}{c}{$\mu$Hz} &
   \multicolumn{1}{c}{$\bar\rho_{\odot}$} &
   \multicolumn{1}{c}{ } &
   \multicolumn{1}{c}{$\mathrm{M}_{\odot}$} &
   \multicolumn{1}{c}{$\mathrm{R}_{\odot}$} &
   \multicolumn{1}{c}{cgs } &
   \multicolumn{1}{c}{cgs} &
   \multicolumn{1}{c}{mas} \\
  \hline 
    KIC3858884$^1$ & $29.0\pm1.0$ & $0.0657\pm0.0021$ & 0.075 & $1.86\pm0.04$ & $3.047\pm0.010$ & $3.740\pm0.012$ & $3.76\pm0.04$ & $1.78\pm0.22$\\
    KIC4544587$^2$ & $74.0\pm1.0$ & $0.414\pm0.039$ & 0.17 & $1.61\pm0.06$ & $1.572\pm0.030$ & $4.252\pm0.033$ & $4.27\pm0.08$ & $1.36\pm0.41$\\
    KIC10661783$^3$ & $39.0\pm1.0$ & $0.1255\pm0.0039$ & 0.20 & $2.100\pm0.028$ & $2.558\pm0.015$ & $3.95\pm0.011$ & $3.95\pm0.04$ & $1.94\pm0.26$\\
    HD172189$^4$ & $19.0\pm1.0$ & $0.0283\pm0.0061$ & 0.28 & $1.78 \pm 0.24$ & $3.98\pm0.11$ & $3.490\pm0.082$ &  $3.53\pm0.06$ & $2.27\pm0.34$\\
    CID100866999$^{5,}$* & $56.0\pm1.0$ & $0.26\pm0.11$ & -- & $1.8\pm0.2$ & $1.9\pm0.2$ & $4.14\pm0.14$ &  -- & -- \\
    CID105906206$^{6,}$* & $20.0\pm2.0$ & $0.02986\pm0.00095$ & 0.15 & $2.25\pm0.04$ & $4.224\pm0.020$ & $3.539\pm0.012$ & $3.52\pm0.10$ & $0.96\pm0.25$\\
    HD159561$^7$ & $38.0\pm1.0$ & $0.124\pm0.021$ & 0.60 & $2.4\pm0.37$ & $2.688\pm0.014$ & $3.960\pm0.072$ & $3.93\pm0.03$ & $67.13\pm1.06$\\
    KIC9851944$^8$ & $26.0\pm1.0$ & $0.0566\pm0.0043$ & 0.29 & $1.79\pm0.07$ & $3.162\pm0.040$ & $3.691\pm0.028$ & $3.75\pm0.25$ & $0.41\pm0.38$\\
    KIC8262223$^9$ & $77.0\pm1.0$ & $0.423\pm0.043$ & 0.11 & $1.96\pm0.06$ & $1.667\pm0.040$ & $4.287\pm0.034$ & $4.23\pm0.10$ & $1.93\pm0.59$\\
    KIC10080943$^{10}$ & $52.0\pm1.0$ & $0.205\pm0.070$ & 0.049 & $1.9\pm0.1$ & $2.10\pm0.20$ & $4.07\pm0.11$ & $4.06\pm0.08$ & $1.06\pm0.28$\\
    HD15082$^{11}$ & $80.0\pm2.0$ & $0.507\pm0.046$ & 0.20 & $1.495\pm0.031$ & $1.434\pm0.034$ & $4.300\pm0.030$ & $4.31\pm0.03$ & $8.51\pm0.24$\\
  \hline
  \hline
   \end{tabular}
 \\ {$^1$\citet{Maceroni2014}; $^2$\citet{Hambleton2013}; $^3$\citet{Lehmann2013}; $^4$\citet{Creevey2009}; $^5$\citet{Chapellier2013}; $^6$\citet{daSilva2014}; $^7$\citet{Monnier2010, Hinkley2011}; $^8$\citet{Guo2016}; $^9$\citet{Guo2017}; $^{10}$\citet{Schmid2015}; $^{11}$\citet{CollierCameron2010}.\\
 *Stars observed by CoRoT have a CoRoT ID of ten numbers. We abbreviate them by not taking into account those left-sided zeros.}
\end{footnotesize}
\end{table*}


\section{The \Dnu-\rhom\ relation}
\label{sec-3}

The analysis of pulsations in \dss\ revealed the presence of periodic patterns in their p-mode frequency spectra \citep[see e.g.][]{garciahernandez2009, garciahernandez2013}. These patterns were also predicted theoretically \citep{Reese2008, Ouazzani2015} and were found to be compatible with a large separation in the low order regime \citep[\Dnu,][]{Suarez2014}. In that paper, it was discovered that the large separation extends to the range $n=[2, 8]$ for non-rotating models. A solid confirmation came from the scaling relation found in GH15 between the observed periodic pattern and the mean density computed for a sample of well-known eclipsing binary stars at different rotation rates. As predicted by oscillation models \citep{Reese2008}, \Dnu\ in \dss\ scales with the stellar mean density independently of its rotation rate.\par

We searched for the large separations of \starj\ and \stark\ from the oscillation spectra publicly available. When necessary, we \emph{cleaned} those spectra in order to remove harmonics and combinations. Likewise, in order to enhance the \Dnu\ pattern in the periodicity analysis, we sought to avoid contamination from g modes, taking only frequencies above $\sim 5$ \cd\ (see GH15). We also considered the recent results in which the observed spacing might be a combination of \Dnu\ and the rotational splitting \citep{Paparo2016sample,Guo2017}. This is the case of \stark, for which the spacing is \Dnu/2 plus the rotational splitting (work in prep.). The methodology employed to find patterns compares the Fourier transforms of the frequencies (assuming all amplitudes equal 1), the histograms of the frequencies differences and the \'{e}chelle diagrams to find the best \Dnu\ candidate, as it is described in GH15.\par

Following GH15, we calculated the mean densities of the new objects of the sample, i.e. considering the Roche model approximation \citep[e.g.][]{Maeder2009}, assuming that the measured radii correspond to the equatorial ones. This allowed us to replace the generally adopted spherical radius by a mean radius, $\bar{R}$, defined as the radius of a sphere with the same volume as the real spheroidal star. Such an approximation was validated through comparisons with more realistic Self-Consistent Field (SCF) models \citep{MacGregor2007}. For computing this mean radius, we need the true rotation velocity of the star, calculated from the projected rotation velocity, the equatorial radius, and the assumption that the inclination of the orbital plane is the same as the rotation axis \citep{Claret1993}.\par

\begin{figure}
	\includegraphics[width=\hsize,clip]{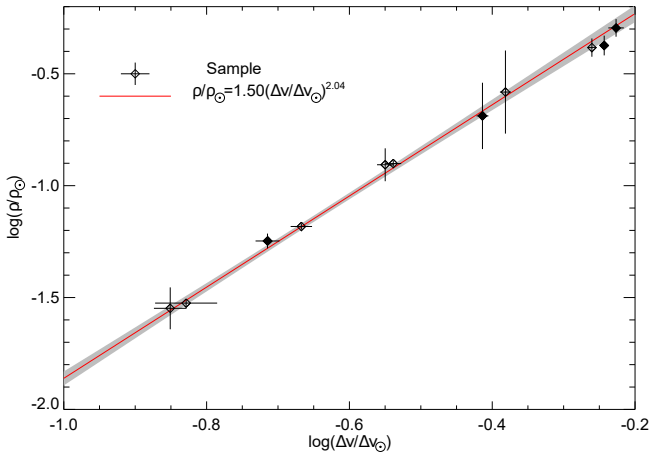}
	\caption{\Dnu\ versus \rhom\ relation with the uncertainties calculated using a Bayesian method. Filled diamonds correspond to the stars added to the GH15 sample.\label{fig-1}}
\end{figure}

We then refined the relation between the mean stellar density and large frequency separation by implementing a Hierarchical Bayesian linear regression. The calculation is implemented using the JAGS package \cite[Just Another Gibbs Sampler,][]{Plummer2003}. A similar application to fit the mass-radius relation of exo-planets can be found in \cite{Wolfgang2015}. The updated relation is:
\begin{equation}
\label{eq1}
\bar\rho/\bar\rho_{\odot}=1.50^{+0.09}_{-0.10}(\Delta\nu/\Delta\nu_{\odot})^{2.04^{+0.04}_{-0.04}},
\end{equation}
 where $\Delta\nu_{\odot}=134.8\ \mu\mathrm{Hz}$ \citep{Kjeldsen2008}. It follows those relations found in previous theoretical works \citep{Suarez2014,Reese2008}. Although this updated relation (see  Fig.~\ref{fig-1}) behaves in a similar way to the one published in GH15, the uncertainties of the fit are one order of magnitude lower for the factor and 50\% lower for the exponent. Again, no dependence with rotation is found in the relation.\par

\section{Asteroseismic determination of the surface gravity}
\label{sec-4}

We computed luminosities and masses through the mass-luminosity relation (MLR) using Gaia's DR1 data. We found accurate parallaxes for 7 of our objects. \starg\ and \stark\ are not present in Gaia's DR1, so we used values from the \Hip database \citep{vanLeeuwen2007} for them. \stard\ belongs to the open cluster IC~4756, whose distance were calculated by \cite{Strassmeier2015}, using \Hip parallaxes of several stars of the cluster. No parallax measurements were found for \stare, so we drop it out from the sample in order to avoid misleading results.\par

The distance modulus relation provided us with the bolometric magnitudes ($\mathrm{M_{bol}}$) that translate into luminosities (with parallaxes measured in mas):

\begin{equation}
\mathrm{M_{bol}}-\mathrm{m_{bol}} = 5 + 5\log\hat{\pi}.
\label{eq2}
\end{equation}

We used visual apparent magnitudes provided by the same bibliographic references as for the parallaxes. To get bolometric magnitudes from visual ones, we needed to apply a bolometric correction (BC) that depends on the effective temperature of the star. However, the BC for A-type stars is usually small \citep[see e.g][]{Hayes1978} so it can be neglected. Indeed, we checked out that this was the case for our sample and duplicated all our calculations with and without taking into account the BC. To compute the corrections, we followed the empirical formulations given by \cite{Torres2010}, obtaining values for BC of the order of 0.01 mag. From $\mathrm{M_{bol}}$, we could obtain the luminosities through the relation: $\mathrm{M_{bol,\odot}}-\mathrm{M_{bol}}=2.5\log\mathrm{L}$, where $\mathrm{M_{bol,\odot}} = 4.74$ and where the luminosity is expressed in solar units. We also took into account the relative luminosities of each star to properly calculate the corresponding flux coming from the pulsating component. The relative luminosities are those obtained from the binary analysis. The only exceptions are \starg\ and \stark, for which we used the total fluxes.\par 
The final step is to use a MLR. Using a sample of well studied and detached Algol-type stars, \cite{Ibanoglu2006} refined the MLR for A-F type stars:

\begin{equation}
\label{eq4}
\mathrm{L}\propto\mathrm{M}_{\hat{\pi}}^{3.92\pm0.05}.
\end{equation}

\begin{figure}
	\includegraphics[width=\hsize,clip]{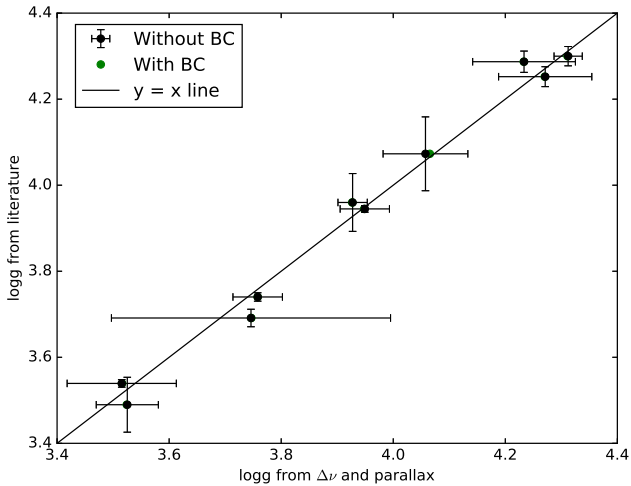}
	\caption{Surface gravities obtained from the analysis carried out in this work versus \logg\ from the binary analysis. Green and black dots represent the calculated values with and without the bolometric correction, respectively. The identity line, $y=x$, is also plotted for reference.\label{fig-2}}
\end{figure}

With the mean density from the \Dnu-\rhom\ relation and the mass from the MLR, we could obtain a radius from the volume of the star assuming spherical symmetry. This radius and the mass allowed us to calculate the surface gravity ($\log g_{\mathrm{\hat{\pi}}}$ in Table~\ref{table1}). The $\log g_{\mathrm{\hat{\pi}}}$ values are remarkably close to those obtained from the binary analysis, which are always within the uncertainties (Table~\ref{table1} and Fig.~\ref{fig-2}). Uncertainties on $\log g_{\mathrm{\hat{\pi}}}$ come from a standard error propagation analysis.\par

We improved the surface gravities from the literature using the provided masses and the mean radius, as defined in Sec.~\ref{sec-3}, to take into account the stellar deformation due to rotation. In general, our calculations give similar results to those in the literature, since the majority of the stars do not move away from sphericity. However, it is important to notice that a unique \logg\ value for rapid rotating stars cannot be defined. This is the case for \starg, for which we followed the same procedure as the others just to show the consistency of the results.\par

We emphasize that masses calculated by this method might not be accurate enough, i. e., within the uncertainties, just \logg. This is because the validity of the MLR relation is for main-sequence stars and some of the stars in our sample are in the H-shell burning phase. Radii are less affected because of the dependency on the cube-root.\par

\section{Discussion}
\label{sec-5}

A few conclusions can be derived immediately from our work. Even when the spherical approximation is used to obtain both the radius and the \logg, the calculated values agreed within the uncertainties with the quantities from the binary analysis. However, the uncertainties coming from the joint analysis of the radial velocity curve and the light curve of an eclipsing binary system are still much lower than those derived using our methodology. This is expected, since masses and radii from the binary analysis are much more precise than those obtained from the MLR and the \Dnu-\rhom\ relation. Indeed, the uncertainty in the exponent of the \Dnu-\rhom\ relation has the largest impact on the accurate determination of \logg. Additionally, precisely determining parallaxes is crucial: the larger the uncertainty on the parallax, the larger the uncertainty on the surface gravity. This explains the large uncertainty for \starh.\par

Nonetheless, we carried out a simple test to check the robustness of the calculated \logg. Instead of deriving masses using parallaxes, we used random values in the range [1, 3] \msol. The maximum dispersion in these new calculated \logg\ was 0.2~dex ($\pm0.1$~dex). This must be considered as an upper limit for the errorbars related to the masses, unless precise masses measurements reduce this uncertainty.

The only value of \logg\ not derived from the binary analysis is that of \starg. The high oblateness of this fast rotator prevented us from getting a unique solution for the surface gravity but rather an average value, which still follows the identity relation. In absence of the actual rotation profile of stars, this new averaged surface gravity might be useful for characterizing rapidly rotating stars. For example, this surface gravity could be compared even to non-rotating models to get an accurate determination of the mass.\par

All the typical error-bars on the surface gravity are of the order of those obtained with high resolution spectroscopy, some of them are even shorter.\par

\section{Conclusions}
\label{sec-con}

In this work, we demonstrated that the analysis of pulsations in \dss\ allows us to accurately obtain some physical parameters of the stars, in particular the mean density and the surface gravity.\par

To achieve these results, we analysed a sample of 10 well studied eclipsing binary systems with a \ds\ component, thus enlarging the sample of GH15. Additionally, we also used the only \ds\ star with a known transiting planet, \stark. For all these systems, precise and model-independent masses and radii were available in the literature. We obtained the large separations for each oscillation spectrum of the pulsating component, following the method by GH15. We correlated the mean densities calculated with the masses and radii with \Dnu. Applying a Hierarchical Bayesian linear regression, we could confirm and reduce the uncertainties on the \Dnu-\rhom\ relation from GH15. Again, we showed that this relation is independent of the rotation velocity of the star.\par

We went then a step further. Using the precise parallaxes given by the Gaia mission, when available, and \Hip\ measurements, we determined the luminosities of the stars. These quantities and the MLR by \cite{Ibanoglu2006} allowed us to estimate masses in an independent way. With this value and the mean density from the \Dnu-\rhom\ relation, we obtained a mean radius from which we calculated surface gravities for the stars of the sample. We checked that the BC has little influence on the calculation of \logg, as expected because it is usually small for A-type stars.\par

We compared surface gravities obtained in this way with those from the literature, obtained from the radial velocity and light curves of the binary systems. We demonstrated that there exists an equivalence between both methods. This fact strengthens the main conclusion of the work: the surface gravity degeneracy can be overcome just analysing the pulsations of \dss\ and using parallaxes measurements.\par

All these results shed light on the understanding of the pulsation spectrum and precise characterisation of \dss, and show that finding \Dnu\ is a major breakthrough to achieve such ambitious goal.\par

\section*{Acknowledgements}
AGH, JCS and AM acknowledge funding support from Spanish public funds for research under project ESP2015-65712-C5-5-R (MINECO/FEDER), and from project RYC-2012-09913 under the `Ram\'on y Cajal' program of the Spanish MINECO. AGH also acknowledges support from Funda\c{c}\~ao para a Ci\^{e}ncia e a Tecnologia (FCT, Portugal) through the fellowship SFRH/BPD/80619/2011. AGH, MJPFGM and JCS acknowledges support from the EC Project SPACEINN (FP7-SPACE-2012-312844). MJPFG also acknowledges partial support from FCT/Portugal through UID/FIS/04434/2013 and CMPETE/FEDER. JPG and RGH acknowledge support from the ``Junta de Andaluc\'{i}a'' local government under project 2012-P12-TIC-2469. SBF acknowledges support from IA (FCT-UID/FIS/04434/2013) to visit Porto, where part of this work was carried out. 


\bibliographystyle{mnras}

\begin{thebibliography}{}
\makeatletter
\relax
\def\mn@urlcharsother{\let\do\@makeother \do\$\do\&\do\#\do\^\do\_\do\%\do\~}
\def\mn@doi{\begingroup\mn@urlcharsother \@ifnextchar [ {\mn@doi@}
  {\mn@doi@[]}}
\def\mn@doi@[#1]#2{\def\@tempa{#1}\ifx\@tempa\@empty \href
  {http://dx.doi.org/#2} {doi:#2}\else \href {http://dx.doi.org/#2} {#1}\fi
  \endgroup}
\def\mn@eprint#1#2{\mn@eprint@#1:#2::\@nil}
\def\mn@eprint@arXiv#1{\href {http://arxiv.org/abs/#1} {{\tt arXiv:#1}}}
\def\mn@eprint@dblp#1{\href {http://dblp.uni-trier.de/rec/bibtex/#1.xml}
  {dblp:#1}}
\def\mn@eprint@#1:#2:#3:#4\@nil{\def\@tempa {#1}\def\@tempb {#2}\def\@tempc
  {#3}\ifx \@tempc \@empty \let \@tempc \@tempb \let \@tempb \@tempa \fi \ifx
  \@tempb \@empty \def\@tempb {arXiv}\fi \@ifundefined
  {mn@eprint@\@tempb}{\@tempb:\@tempc}{\expandafter \expandafter \csname
  mn@eprint@\@tempb\endcsname \expandafter{\@tempc}}}

\bibitem[\protect\citeauthoryear{Baglin, Auvergne, Barge, Deleuil, Catala,
  Michel, Weiss  \& {The COROT Team}}{Baglin et~al.}{2006}]{corot}
Baglin A.,  Auvergne M.,  Barge P.,  Deleuil M.,  Catala C.,  Michel E.,  Weiss
  W.,   {The COROT Team} 2006, in Fridlund M.,  Baglin A.,  Lochard J.,
  Conroy L.,  eds,  ESA Special Publication Vol. 1306, ESA Special Publication.
  p.~33

\bibitem[\protect\citeauthoryear{Barcel{\'{o}}~Forteza, Roca~Cort{\'{e}}s,
  Garc{\'{i}}a~Hernandez  \& Garc{\'{i}}a}{Barcel{\'{o}}~Forteza
  et~al.}{2017}]{Sebastia2017}
Barcel{\'{o}}~Forteza S.,  Roca~Cort{\'{e}}s T.,  Garc{\'{i}}a~Hernandez A.,
  Garc{\'{i}}a R.~A.,  2017, Astronomy {\&} Astrophysics, 601, A57

\bibitem[\protect\citeauthoryear{Cameron et~al.,}{Cameron
  et~al.}{2010}]{CollierCameron2010}
Cameron A.~C.,  et~al., 2010, Monthly Notices of the Royal Astronomical
  Society, 407, 507

\bibitem[\protect\citeauthoryear{Chapellier \& Mathias}{Chapellier \&
  Mathias}{2013}]{Chapellier2013}
Chapellier E.,  Mathias P.,  2013, Astronomy {\&} Astrophysics, 556, A87

\bibitem[\protect\citeauthoryear{Claret \& Gimenez}{Claret \&
  Gimenez}{1993}]{Claret1993}
Claret A.,  Gimenez A.,  1993, Astronomy {\&} Astrophysics, 277, 487

\bibitem[\protect\citeauthoryear{Creevey et~al.,}{Creevey
  et~al.}{2009}]{Creevey2009}
Creevey O.~L.,  et~al., 2009, Astronomy {\&} Astrophysics, 507, 901

\bibitem[\protect\citeauthoryear{Gaia~Collaboration et~al.,}{Gaia~Collaboration
  et~al.}{2016a}]{gaiamission}
Gaia~Collaboration G.,  et~al., 2016a, Astronomy {\&} Astrophysics, 595, A1

\bibitem[\protect\citeauthoryear{Gaia~Collaboration et~al.,}{Gaia~Collaboration
  et~al.}{2016b}]{gaiaDR1}
Gaia~Collaboration G.,  et~al., 2016b, Astronomy {\&} Astrophysics, 595, A2

\bibitem[\protect\citeauthoryear{Garc{\'{i}}a~Hern{\'{a}}ndez
  et~al.,}{Garc{\'{i}}a~Hern{\'{a}}ndez et~al.}{2009}]{garciahernandez2009}
Garc{\'{i}}a~Hern{\'{a}}ndez A.,  et~al., 2009, Astronomy {\&} Astrophysics,
  506, 79

\bibitem[\protect\citeauthoryear{Garc{\'{i}}a~Hern{\'{a}}ndez
  et~al.,}{Garc{\'{i}}a~Hern{\'{a}}ndez et~al.}{2013}]{garciahernandez2013}
Garc{\'{i}}a~Hern{\'{a}}ndez A.,  et~al., 2013, Astronomy {\&} Astrophysics,
  559, A63

\bibitem[\protect\citeauthoryear{Garc{\'{i}}a~Hern{\'{a}}ndez,
  Mart{\'{i}}n-Ruiz, Monteiro, Su{\'{a}}rez, Reese, Pascual-Granado  \&
  Garrido}{Garc{\'{i}}a~Hern{\'{a}}ndez et~al.}{2015}]{gh2015}
Garc{\'{i}}a~Hern{\'{a}}ndez A.,  Mart{\'{i}}n-Ruiz S.,  Monteiro M. J. P.
  F.~G.,  Su{\'{a}}rez J.~C.,  Reese D.~R.,  Pascual-Granado J.,   Garrido R.,
  2015, The Astrophysical Journal Letters, 811, L29

\bibitem[\protect\citeauthoryear{Goupil et~al.,}{Goupil
  et~al.}{2005}]{Goupil2005}
Goupil M.~J.,  et~al., 2005, Journal of Astrophysics {\&} Astronomy, 26, 249

\bibitem[\protect\citeauthoryear{Guo, Gies, Matson  \&
  Garc{\'{i}}a~Hern{\'{a}}ndez}{Guo et~al.}{2016}]{Guo2016}
Guo Z.,  Gies D.~R.,  Matson R.~A.,   Garc{\'{i}}a~Hern{\'{a}}ndez A.,  2016,
  The Astrophysical Journal, 826, 69

\bibitem[\protect\citeauthoryear{Guo, Gies, Matson,
  Garc{\'{i}}a~Hern{\'{a}}ndez, Han  \& Chen}{Guo et~al.}{2017}]{Guo2017}
Guo Z.,  Gies D.~R.,  Matson R.~A.,  Garc{\'{i}}a~Hern{\'{a}}ndez A.,  Han Z.,
   Chen X.,  2017, The Astrophysical Journal, 837, 114

\bibitem[\protect\citeauthoryear{Hambleton et~al.,}{Hambleton
  et~al.}{2013}]{Hambleton2013}
Hambleton K.~M.,  et~al., 2013, Monthly Notices of the Royal Astronomical
  Society, 434, 925

\bibitem[\protect\citeauthoryear{Hayes}{Hayes}{1978}]{Hayes1978}
Hayes D.~S.,  1978, in {IAU} ed.,  Vol. 80, The HR diagram - The 100th
  anniversary of Henry Norris Russell. D. Reidel Publishing Co., pp 65--75

\bibitem[\protect\citeauthoryear{Hinkley et~al.,}{Hinkley
  et~al.}{2010}]{Hinkley2011}
Hinkley S.,  et~al., 2010, The Astrophysical Journal, 726, 104

\bibitem[\protect\citeauthoryear{Ibanoglu, Soydugan, Soydugan  \&
  Dervisoglu}{Ibanoglu et~al.}{2006}]{Ibanoglu2006}
Ibanoglu C.,  Soydugan F.,  Soydugan E.,   Dervisoglu A.,  2006, Monthly
  Notices of the Royal Astronomical Society, 373, 435

\bibitem[\protect\citeauthoryear{Kjeldsen, Bedding  \&
  Christensen-Dalsgaard}{Kjeldsen et~al.}{2008}]{Kjeldsen2008}
Kjeldsen H.,  Bedding T.~R.,   Christensen-Dalsgaard J.,  2008, The
  Astrophysical Journal Letters, 683, L175

\bibitem[\protect\citeauthoryear{Koch et~al.,}{Koch et~al.}{2010}]{kepler}
Koch D.~G.,  et~al., 2010, The Astrophysical Journal Letters, 713, L79

\bibitem[\protect\citeauthoryear{Lehmann, Southworth, Tkachenko  \&
  Pavlovski}{Lehmann et~al.}{2013}]{Lehmann2013}
Lehmann H.,  Southworth J.,  Tkachenko A.,   Pavlovski K.,  2013, Astronomy
  {\&} Astrophysics, 557, A79

\bibitem[\protect\citeauthoryear{MacGregor, Jackson, Skumanich  \&
  Metcalfe}{MacGregor et~al.}{2007}]{MacGregor2007}
MacGregor K.~B.,  Jackson S.,  Skumanich A.,   Metcalfe T.~S.,  2007, The
  Astrophysical Journal, 663, 560

\bibitem[\protect\citeauthoryear{Maceroni et~al.,}{Maceroni
  et~al.}{2014}]{Maceroni2014}
Maceroni C.,  et~al., 2014, Astronomy {\&} Astrophysics, 563, A59

\bibitem[\protect\citeauthoryear{Maeder \& {Andr{\'{e}}}}{Maeder \&
  {Andr{\'{e}}}}{2009}]{Maeder2009}
Maeder A.,  {Andr{\'{e}}} 2009, {Physics, Formation and Evolution of Rotating
  Stars}.
Astronomy and Astrophysics Library, Springer Berlin Heidelberg, Berlin,
  Heidelberg

\bibitem[\protect\citeauthoryear{Michel et~al.,}{Michel
  et~al.}{2017}]{Michel2017}
Michel E.,  et~al., 2017, ArXiv

\bibitem[\protect\citeauthoryear{Monnier, Townsend, Che, Zhao, Kallinger,
  Matthews  \& Moffat}{Monnier et~al.}{2010}]{Monnier2010}
Monnier J.~D.,  Townsend R. H.~D.,  Che X.,  Zhao M.,  Kallinger T.,  Matthews
  J.,   Moffat A. F.~J.,  2010, The Astrophysical Journal, 725, 1192

\bibitem[\protect\citeauthoryear{Ouazzani, Dupret  \& Reese}{Ouazzani
  et~al.}{2012}]{Ouazzani2012}
Ouazzani R.-M.,  Dupret M.-A.,   Reese D.~R.,  2012, Astronomy {\&}
  Astrophysics, 547, A75

\bibitem[\protect\citeauthoryear{Ouazzani, Roxburgh  \& Dupret}{Ouazzani
  et~al.}{2015}]{Ouazzani2015}
Ouazzani R.-M.,  Roxburgh I.~W.,   Dupret M.-A.,  2015, Astronomy {\&}
  Astrophysics, 579, A116

\bibitem[\protect\citeauthoryear{Papar{\'{o}}, Benk{\H{o}}, Hareter  \&
  Guzik}{Papar{\'{o}} et~al.}{2016}]{Paparo2016sample}
Papar{\'{o}} M.,  Benk{\H{o}} J.~M.,  Hareter M.,   Guzik J.~A.,  2016, The
  Astrophysical Journal Supplement Series, Volume 224, 224, 41

\bibitem[\protect\citeauthoryear{Pascual-Granado, Garrido  \&
  Su{\'{a}}rez}{Pascual-Granado et~al.}{2015}]{Javier2015}
Pascual-Granado J.,  Garrido R.,   Su{\'{a}}rez J.~C.,  2015, Astronomy {\&}
  Astrophysics, 581, A89

\bibitem[\protect\citeauthoryear{Plummer}{Plummer}{2003}]{Plummer2003}
Plummer M.,  2003, in Hornik K.,  Leisch F.,  Zeileis A.,   Plummer M.,  eds,
  Proceedings of the 3rd International Workshop on Distributed Statistical
  Computing.

\bibitem[\protect\citeauthoryear{Poretti et~al.,}{Poretti
  et~al.}{2009}]{poretti2009}
Poretti E.,  et~al., 2009, Astronomy {\&} Astrophysics, 506, 85

\bibitem[\protect\citeauthoryear{Reese, Ligni{\`{e}}res  \& Rieutord}{Reese
  et~al.}{2008}]{Reese2008}
Reese D.,  Ligni{\`{e}}res F.,   Rieutord M.,  2008, Astronomy {\&}
  Astrophysics, 481, 449

\bibitem[\protect\citeauthoryear{Reese, Prat, Barban, van'~t Veer-Menneret  \&
  MacGregor}{Reese et~al.}{2013}]{Reese2013}
Reese D.~R.,  Prat V.,  Barban C.,  van'~t Veer-Menneret C.,   MacGregor K.~B.,
   2013, Astronomy {\&} Astrophysics, 550, 77

\bibitem[\protect\citeauthoryear{Reese, Ligni{\`{e}}res, Ballot, Dupret,
  Barban, van~’t Veer-Menneret  \& MacGregor}{Reese et~al.}{2017}]{Reese2017}
Reese D.~R.,  Ligni{\`{e}}res F.,  Ballot J.,  Dupret M.-A.,  Barban C.,
  van~’t Veer-Menneret C.,   MacGregor K.~B.,  2017, Astronomy {\&}
  Astrophysics, 601, A130

\bibitem[\protect\citeauthoryear{Rieutord, Lara  \& Putigny}{Rieutord
  et~al.}{2016}]{Rieutord2016}
Rieutord M.,  Lara F.~E.,   Putigny B.,  2016, Journal of Computational
  Physics, 318, 277

\bibitem[\protect\citeauthoryear{Royer, Zorec  \& G{\'{o}}mez}{Royer
  et~al.}{2007}]{Royer2007}
Royer F.,  Zorec J.,   G{\'{o}}mez A.~E.,  2007, Astronomy {\&} Astrophysics,
  463, 671

\bibitem[\protect\citeauthoryear{Schmid et~al.,}{Schmid
  et~al.}{2015}]{Schmid2015}
Schmid V.~S.,  et~al., 2015, Astronomy {\&} Astrophysics, 584, A35

\bibitem[\protect\citeauthoryear{Strassmeier, Weingrill, Granzer, Bihain, Weber
   \& Barnes}{Strassmeier et~al.}{2015}]{Strassmeier2015}
Strassmeier K.~G.,  Weingrill J.,  Granzer T.,  Bihain G.,  Weber M.,   Barnes
  S.~A.,  2015, Astronomy {\&} Astrophysics, 580, A66

\bibitem[\protect\citeauthoryear{Su{\'{a}}rez, Garc{\'{i}}a~Hern{\'{a}}ndez,
  Moya, Rodrigo, Solano, Garrido  \& Rod{\'{o}}n}{Su{\'{a}}rez
  et~al.}{2014}]{Suarez2014}
Su{\'{a}}rez J.~C.,  Garc{\'{i}}a~Hern{\'{a}}ndez A.,  Moya A.,  Rodrigo C.,
  Solano E.,  Garrido R.,   Rod{\'{o}}n J.~R.,  2014, Astronomy {\&}
  Astrophysics, 563, 11

\bibitem[\protect\citeauthoryear{Torres}{Torres}{2010}]{Torres2010}
Torres G.,  2010, The Astronomical Journal, 140, 1158

\bibitem[\protect\citeauthoryear{Walker et~al.,}{Walker et~al.}{2003}]{most}
Walker G.,  et~al., 2003, Publications of the Astronomical Society of the
  Pacific, 115, 1023

\bibitem[\protect\citeauthoryear{Wolfgang, Rogers  \& Ford}{Wolfgang
  et~al.}{2015}]{Wolfgang2015}
Wolfgang A.,  Rogers L.~A.,   Ford E.~B.,  2015, in IAU General Assembly,
  Meeting {\#}29.

\bibitem[\protect\citeauthoryear{da Silva, Maceroni, Gandolfi, Lehmann  \&
  Hatzes}{da~Silva et~al.}{2014}]{daSilva2014}
da Silva R.,  Maceroni C.,  Gandolfi D.,  Lehmann H.,   Hatzes A.~P.,  2014,
  Astronomy {\&} Astrophysics, 565, A55

\bibitem[\protect\citeauthoryear{van Leeuwen \& {Floor}}{van Leeuwen \&
  {Floor}}{2010}]{vanLeeuwen2007}
van Leeuwen F.,  {Floor} 2010, Space Science Reviews, 151, 209

\makeatother
\end{thebibliography}

\end{document}